\documentclass[11pt,document,nofootinbib,superscriptaddress,onecolumn,preprintnumbers,balancelastpage]{article}
\pdfoutput=1
\usepackage{jheppub}
\usepackage{graphicx}
\usepackage{epstopdf}
\usepackage{dcolumn}
\usepackage{bm}
\usepackage{hyperref}
\usepackage{booktabs}
\usepackage[dvipsnames]{xcolor}
\usepackage{amsmath}
\usepackage{cancel}
\usepackage{xpatch}
\usepackage{mathtools}
\usepackage{multirow}
\usepackage{bbold}
\usepackage{soul}

\makeatletter
\g@addto@macro\bfseries{\boldmath}
\makeatother

\newcommand{\GeV}{\textrm{ GeV}}

\newcommand{\vev}[1]{\langle #1 \rangle}

\title{
Naturally astrophobic QCD axion
}
\author[1]{Marcin Badziak}
\author[2,3,4,5]{Keisuke Harigaya}
\affiliation[1]{Institute of Theoretical Physics, Faculty of Physics, University of Warsaw, ul. Pasteura 5, PL-02-093 Warsaw, Poland }
\affiliation[2]{Department of Physics, University of Chicago, Chicago, IL 60637, USA}
\affiliation[3]{Enrico Fermi Institute, University of Chicago, Chicago, IL 60637, USA}
\affiliation[4]{Kavli Institute for Cosmological Physics, University of Chicago, Chicago, IL 60637, USA}
\affiliation[5]{Kavli Institute for the Physics and Mathematics of the Universe (WPI),
The University of Tokyo Institutes for Advanced Study,
The University of Tokyo, Kashiwa, Chiba 277-8583, Japan}

\abstract{
We present a QCD axion model where the couplings of the axion to nucleons, electrons, and muons are naturally suppressed because of the appropriate choice of the Peccei-Quinn charges of the Standard Model fermions. We reexamine next-to-leading order corrections to the couplings of the axion with nucleons and photons and show that  the axion decay constant may be as small as $10^7$ GeV. It is also possible to suppress the coupling with the photon so that the decay constant is even smaller and minimal axiogenesis works. In this scenario, the axion has a mass above 1 eV and may be directly detected via absorption of axion dark matter. Flavor-violating axion couplings are generically predicted in our model, but we show that they may be naturally and sufficiently suppressed. We discuss the implications of the hints for anomalous cooling in several stellar environments to our model.
}

\begin{document}

\maketitle


\section{Introduction}

The strong CP problem may be solved by the Peccei-Quinn (PQ) mechanism~\cite{Peccei:1977hh,Peccei:1977ur} that predicts a hypothetical particle called the QCD axion~\cite{Weinberg:1977ma,Wilczek:1977pj}. The axion is also a good dark-matter (DM) candidate~\cite{Preskill:1982cy, Dine:1982ah,Abbott:1982af}.
In typical QCD axion models, phenomenologically viable axion masses are much below the eV scale. The strongest upper bounds on the QCD axion mass come from astrophysics. While the structure of axion couplings varies among different models, in most models an axion-nucleon coupling is present, leading to an upper bound from the observations of the neutrino burst in SN1987A~\cite{Ellis:1987pk,Mayle:1987as,Turner:1987by,Raffelt:1987yt,Chang:2018rso,Carenza:2019pxu} and the cooling of neutron stars~\cite{Iwamoto:1984ir,Iwamoto:1992jp,Page:2010aw,Shternin:2010qi,Sedrakian:2015krq,Hamaguchi:2018oqw,Sedrakian:2018kdm,Leinson:2021ety,Buschmann:2021juv}. For example, in the minimal KSVZ~\cite{Kim:1979if,Shifman:1979if} and DFSZ models~\cite{Zhitnitsky:1980tq,Dine:1981rt}, these observations lead to an upper bound on the axion mass of $\mathcal{O}(10^{-2})$~eV or, equivalently, to a lower bound on the axion decay constant of $f_a\gtrsim\mathcal{O}(10^9)$~GeV.

There are theoretical and phenomenological motivations to consider values of $f_a$ much smaller than that allowed in minimal QCD axion models.
On the theoretical side, the axion quality problem is relaxed for smaller decay constants~\cite{Holman:1992us,Barr:1992qq,Kamionkowski:1992mf,Dine:1992vx}. On the phenomenological side, low $f_a$ is preferred to explain  the observed baryon asymmetry simultaneously with the observed DM abundance in the axiogenesis scenario~\cite{Co:2019wyp} in which axion DM is produced via the kinetic misalignment mechanism~\cite{Co:2019jts}. Such a scenario predicts the QCD axion mass around the eV scale that can be detected using optical haloscopes~\cite{Arvanitaki:2017nhi,Baryakhtar:2018doz}, which are based on absorption of DM, such as the LAMPOST experiment~\cite{Chiles:2021gxk}.

All of the above motivate construction of QCD axion models in which axion-nucleon couplings are strongly suppressed. Such models have been proposed and dubbed astrophobic axion models in~\cite{DiLuzio:2017ogq}. Astrophobic axion models to deserve their name should also avoid astrophysical constraints on the other axion couplings, especially the axion-electron coupling that also leads to a lower bound on $f_a\sim\mathcal{O}(10^9)$~GeV from White Dwarfs (WDs) unless the axion does not couple to electrons at tree level~\cite{Raffelt:1985nj,Blinnikov:1994eoa,MillerBertolami:2014rka}. Astrophobic axion models proposed in~\cite{DiLuzio:2017ogq} generalize the DFSZ model by introducing flavor non-universal PQ charges that allow for very small axion-nucleon couplings by fine-tuning. In such models axion-lepton couplings are generically not suppressed but the bound on the axion-electron coupling from WDs can be satisfied at the cost of additional tuning of the model parameters. A generalization of this setup with three Higgs doublets was proposed in \cite{Bjorkeroth:2019jtx} in which simultaneous suppression of the axion couplings to nucleons and electron is achieved by a single tuning of parameters.

The goal of this paper is to construct natural astrophobic axion models in which the suppression of astrophysically relevant axion couplings does not require tuning of parameters but stems from the PQ charge assignment of the Standard Model (SM) fermions.
Indeed, if the up and down quarks have PQ charges of $2$ and $1$, respectively, and there is no QCD anomaly of the PQ symmetry beyond that from the up and down quarks, axion-nucleon couplings are suppressed. Vanishing PQ charges of the electron and muon also ensure that the axion does not couple to them at the tree level. 

Unlike KSVZ and DFSZ axions, astrophobic axions are generically flavor violating. The axion-down-strange coupling, which is strongly bounded from kaon decay~\cite{E949:2007xyy,MartinCamalich:2020dfe}, can be naturally suppressed if the PQ charges of down and strange quarks are the same. Instead, special flavor symmetry can suppress the coupling without assuming the same charges of the down and strange quarks. We also study other flavor-violating couplings and show that they are also sufficiently small.

It is also possible to  suppress the axion-photon coupling. This accidentally occurs if the electromagnetic anomaly coefficient of the PQ symmetry, $E$, is twice larger than that of QCD anomaly, $N$, as pointed out in \cite{Kaplan:1985dv}. The decay constant may be then below $10^7$ GeV and as small as $10^6$ GeV. The minimal axiogenesis, which suffers from the overproduction of axion DM by kinetic misalignment in the KSVZ and DFSZ models, becomes successful.

We also discuss the implications of the hints for anomalous cooling in several stellar environments~\cite{Raffelt:2011ft,Giannotti:2015kwo,Giannotti:2017hny} that can be explained by the axion-electron and/or photon couplings. If the electron coupling is generated by the quantum correction from the axion-gluon or axion-$W$ coupling, the cooling anomalies can be explained if $E/N=2$. It is also possible to have a small tree-level axion-electron coupling, for which the best-fit value can be obtained even if $E/N\neq2$.

We present UV completions of natural astrophobic axion models that include vector-like fermions or Higgs doublets, which can be thought of as generalized KSVZ and DFSZ models, respectively.

The rest of the article is organized as follows. In Sec.~\ref{sec:EFT}, we discuss the coupling of the axion with SM particles at an effective field theory level and show that the decay constant may be naturally $\mathcal{O}(10^{6-7})$ GeV. In Sec.~\ref{sec:UV}, we discuss UV completions of the setup. Stellar-cooling anomalies and axiogenesis are discussed in Secs.~\ref{sec:cooling} and \ref{sec:axiogenesis}, respectively. Sec.~\ref{sec:summary} gives summary and discussion.

\section{Axion couplings}
\label{sec:EFT}

In this section, we discuss the coupling of the axion to the SM particles in the effective field theory with the SM particles and the axion. UV completions are discussed in Sec.~\ref{sec:UV}. We consider a class of models such that the SM fermions are charged under the PQ symmetry and the phases of the Yukawa couplings depend on the axion field. Unlike the models in Refs.~\cite{Ema:2016ops,Calibbi:2016hwq}, we do not attempt to explain the flavor hierarchy solely by the PQ symmetry. Rather, we only require that the model does not require any unnatural structure. We denote the PQ charge of a left-handed Weyl fermion $f$ as $Q_f$.

Without much loss of generality, we consider the case where only the right-handed fermions $\bar{u}$, $\bar{c}$, $\bar{t}$, $\bar{d}$, $\bar{s}$, $\bar{b}$, $\bar{e}$, $\bar{\mu}$, and $\bar{\tau}$  are charged under the PQ symmetry. In the limit where the Yukawa couplings are diagonal, even if some of the left-handed fermions are PQ-charged, we may take a linear combination of the PQ symmetry and baryon and lepton symmetry of each generation to make the left-handed fermions neutral under the PQ symmetry. The generality is lost for flavor-violating axion couplings. As we will see, however, PQ-charged left-handed fermions typically lead to larger flavor-violating coupling unless the PQ charge is generation independent, for which the PQ charge of left-handed fermions can be removed by combining it with flavor-universal baryon or lepton symmetry. Thus, models where only right-handed fermions are PQ-charged are the most conservative ones.

\subsection{Quark and nucleon couplings}
If $Q_{\bar{u}}/Q_{\bar{d}}=2$ and there is no extra QCD anomaly beyond that from $\bar{u}$ and $\bar{d}$, the axion-nucleon coupling is suppressed. This can be most easily seen in the basis where the up and down quark masses depend on the axion field. In this basis, the kinetic mixing between the axion and the pion is absent. Also,  since $m_u/m_d \simeq 0.5$,  the axion-dependent quark mass is isospin singlet to the leading order in $1/f_a$, and the axion-pion mass mixing is suppressed.

Let us explicitly see the suppression of the nucleon coupling, including quantum corrections. We assign $Q_{\bar{u}}=2$ and $Q_{\bar{d}}=1$.
We also assign $Q_{\bar{t}}=0$. Otherwise, one-loop RG correction from the axion-top coupling generates the axion-up and -down coupling that contributes to the axion-nucleon coupling~\cite{Choi:2017gpf,Choi:2021kuy,Bauer:2020jbp}, and the axion is no longer astrophobic unless that coupling is canceled by fine-tuning~\cite{DiLuzio:2022tyc}.

\begin{table}[!t]
    \centering
    \begin{tabular}{|c|c|c|} \hline
       $g_A^u - g_A^d$  & \multicolumn{2}{c|}{1.2723(23)}    \\ \hline \hline 
         & $N_f=2+1+1$ & $N_f = 2+1$ \\ \hline
       $g_A^u + g_A^d$  & 0.34(5)  & 0.44(4) \\
       $\delta_s$  & 0.059(8) & 0.044(9)  \\
       $\delta_c$  & 0.0065(39) & 0.0092(39)   \\
       $\delta_b$ &  0.0045(12)&  0.0063(15)  \\
       $z= m_u/m_d$ & 0.465(24) & 0.485(19) \\
       $w=m_u/m_s$  & 0.023(1) & 0.024(1) \\
       \hline
    \end{tabular}
    \caption{Numerical values of the constants that determine the axion-nucleon coupling.}
    \label{tab:coeff}
\end{table}

When the QCD anomaly comes only from the up and down quarks, after removing the axion field from the fermion mass terms by chiral rotation, the axion-fermion couplings at a UV scale are
\begin{align}
\label{eq:quarkcoup}
\frac{\partial_\mu a}{f_a} \sum_f c_f f^\dag \bar{\sigma}^\mu f,~~ c_f=- \frac{Q_f}{3}.
\end{align}
Here the factor of $3$ comes from the QCD anomaly coefficient of $3$.
Below the electroweak symmetry breaking scale, it is convenient to write the interaction in terms of the axial current of Dirac fields $\psi$,
\begin{align}
\label{eq:coupling_dirac}
\frac{\partial a}{2f_a} \sum_\psi C_\psi \bar{\psi} \gamma^\mu \gamma^5 \psi,~~ C_u = - c_{\bar{u}},~~ C_d = - c_{\bar{d}},~~C_e = - c_{\bar{e}}, \cdots.
\end{align}
Following the computation in~\cite{GrillidiCortona:2015jxo,Vonk:2020zfh}, we find that the axion-nucleon couplings are given by
\begin{align}
\label{eq:axion-N}
&\frac{\partial_\mu a}{2f_a} \sum_{N=p,n}C_N \bar{N} \gamma^\mu \gamma_5 N,\\
C_{p} - C_n = & \left(g_{A}^u - g_{A}^d\right) \left(C_u-C_d- \frac{1-z}{1+z+w}\right), \nonumber \\
C_{p} + C_n = & \left(g_{A}^u + g_{A}^d\right) \left(0.95\left(C_u + C_d\right) + 0.05 -\frac{1+z}{1+z+w}\right) 
- 2 \delta, \nonumber \\
\delta = & \sum_{i=s,c,b}{\delta_i}C_i + \frac{m_\pi^2}{m_{\eta'}^2} \frac{f_\pi}{m_{N}} \frac{\sqrt{6}z}{(1+z)^2}\times G.\nonumber
\end{align}
Here $N$ is a Dirac field.
The values of $g_A^u-g_A^d$, $g_A^u+g_A^d$, $\delta_i$, $z=m_u/m_d$, and $w = m_u/m_s$ are shown in Table~\ref{tab:coeff}.
Here we used the up-to-date lattice data for $N_f=2+1+1$ and $2+1$, with $N_f$ being the number of flavors,  summarized in~\cite{FlavourLatticeAveragingGroupFLAG:2021npn}.
Since the errors of two cases are comparable, we use the average of them. 
The RGE running is computed from $10^7$~GeV, although the change of the coefficients by using a different UV scale is negligible in comparison with the errors.  
The uncertainties in the coefficients $\delta_{s,c,b}$ are dominated by those of the nucleon matrix element of the strange axial current $(g_A^s)$, that of charm $(g_A^c)$, the RG evolution of the axion-quark couplings and the nucleon matrix element of isospin-singlet up and down axial current $(g_A^{u+d})$, respectively.  
Since $C_u=2/3$, $C_d=1/3$, and $z \simeq 0.5$, the axion-nucleon coupling is indeed suppressed.
The astrophysical constraints are often expressed in terms of dimensionless combinations $g_{aNN}\equiv C_Nm_N/f_a$.

We included $\mathcal{O}(m_u/m_s)$ effect and an $\mathcal{O}(m_\pi^2/m_{\eta'}^2)$ term in $\delta$ that are not included in the previous literature on astrophobic axions. These are usually negligible, but are important for out setup where the leading-order axion-nucleon coupling naturally vanishes.
The last term in $\delta$ originates from the axion-dependent quark mass in the basis where the axion-pion and axion-eta mixing from the quark mass is removed. To the reading order in $1/f_a$, this becomes the coupling of the axion to the $SU(3)$-singlet pseudo-scalar quark bilinear. We may estimate the induced axion-nucleon coupling from the axion-eta' mixing
\begin{align}
\theta_{a\eta'} \sim \frac{m_\pi^2}{m_{\eta'}^2} \frac{\sqrt{6}z}{(1+z)^2} \frac{f_\pi}{f_a} 
\end{align}
and the zero-momentum limit of the eta'-nucleon coupling $g_{\eta'NN}$. The constant $G$ in Eq.~(\ref{eq:axion-N}) is expected to be ${\cal O}(g_{\eta'NN})$. Theoretical estimations based on the sum rule and the Goldberger-Treiman relation find $g_{\eta'NN}= {\cal O}(1)$~\cite{Feldmann:1999uf,Singh:2018yvt}.

\begin{table}[!t]
    \centering
    \begin{tabular}{|c|c|c|} \hline
      & $\bar{u}$ & $\bar{d}$  \\ \hline
      $Q_f$ & $2$  & $1$  \\ \hline
    \end{tabular}
    \caption{The PQ charges $Q_f$ of the SM fermions in the minimal model. Fermions other than $\bar{u}$ and $\bar{d}$ are PQ neutral. There should be no QCD anomaly of the PQ symmetry beyond that from the up and down quarks.}
    \label{tab:minimal}
\end{table}

\begin{figure}[!t]
\centering
\includegraphics[width=0.49\linewidth]{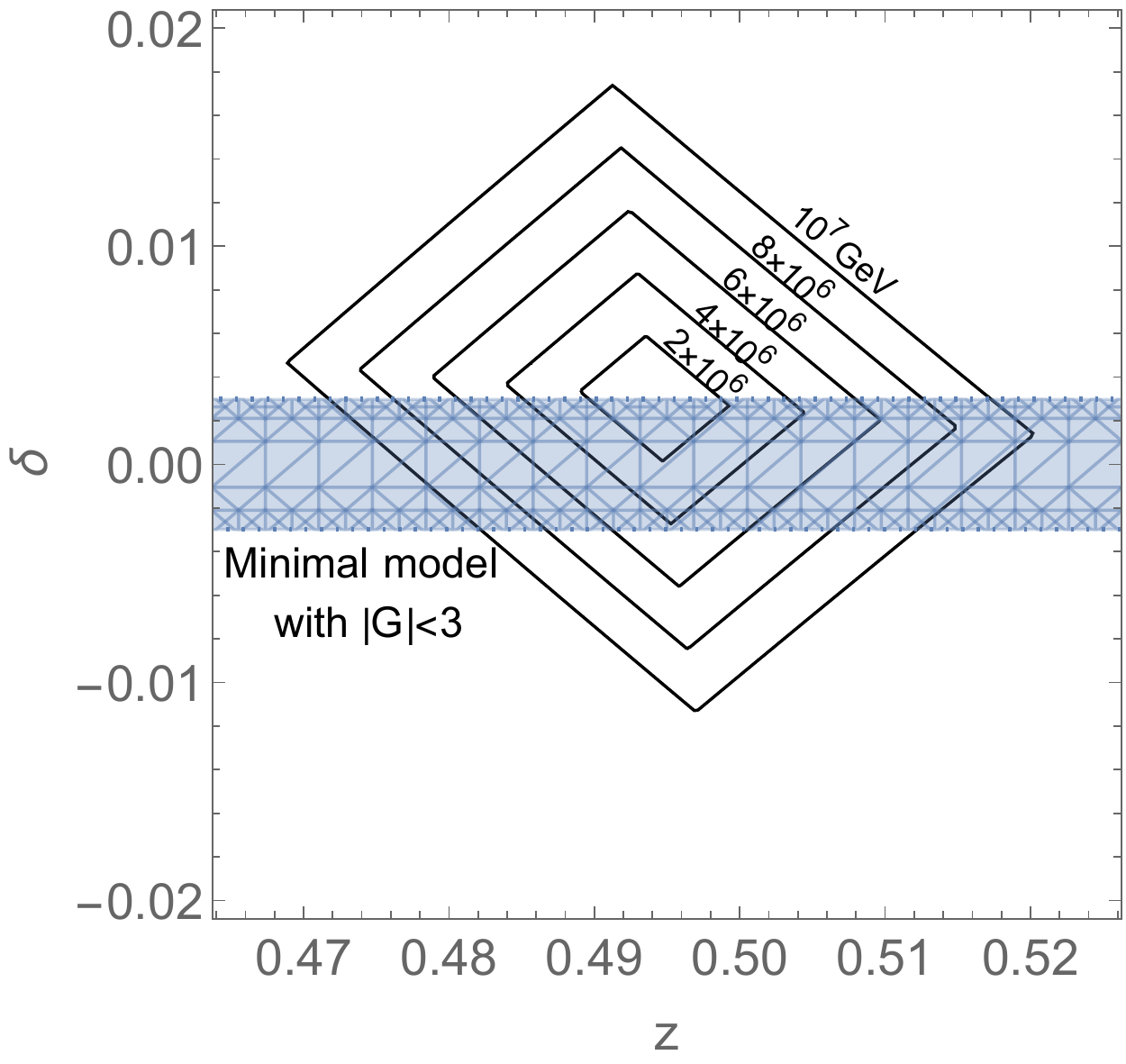}
	\caption{The lower bound on the decay constant $f_a$ from the cooling of neutron stars for given $z$ and $\delta$. The expected value of $\delta$ in the minimal model in Table~\ref{tab:minimal} is shown by the blue band.}
	\label{fig:nucleon_minimal}	
\end{figure}

In Fig.~\ref{fig:nucleon_minimal}, we show the lower bound on $f_a$ from the cooling of neutron stars, $g_{app} < 1.5 \times 10^{-9}$ and $g_{ann} < 1.2 \times 10^{-9}$~\cite{Buschmann:2021juv}, as a function of $z$ and $\delta$.
In the minimal model, only $\bar{u}$ and $\bar{d}$ are PQ charged, as shown in Table~\ref{tab:minimal}, so $\delta$ is determined by the last term. The blue band shows the expectation for such a case assuming $|G|<3$. For the allowed range of $z$, $f_a$ may be below $10^7$ GeV, or even below few $10^6$ GeV.

\begin{figure}[!t]
 \centering
  \includegraphics[width=0.49\linewidth]{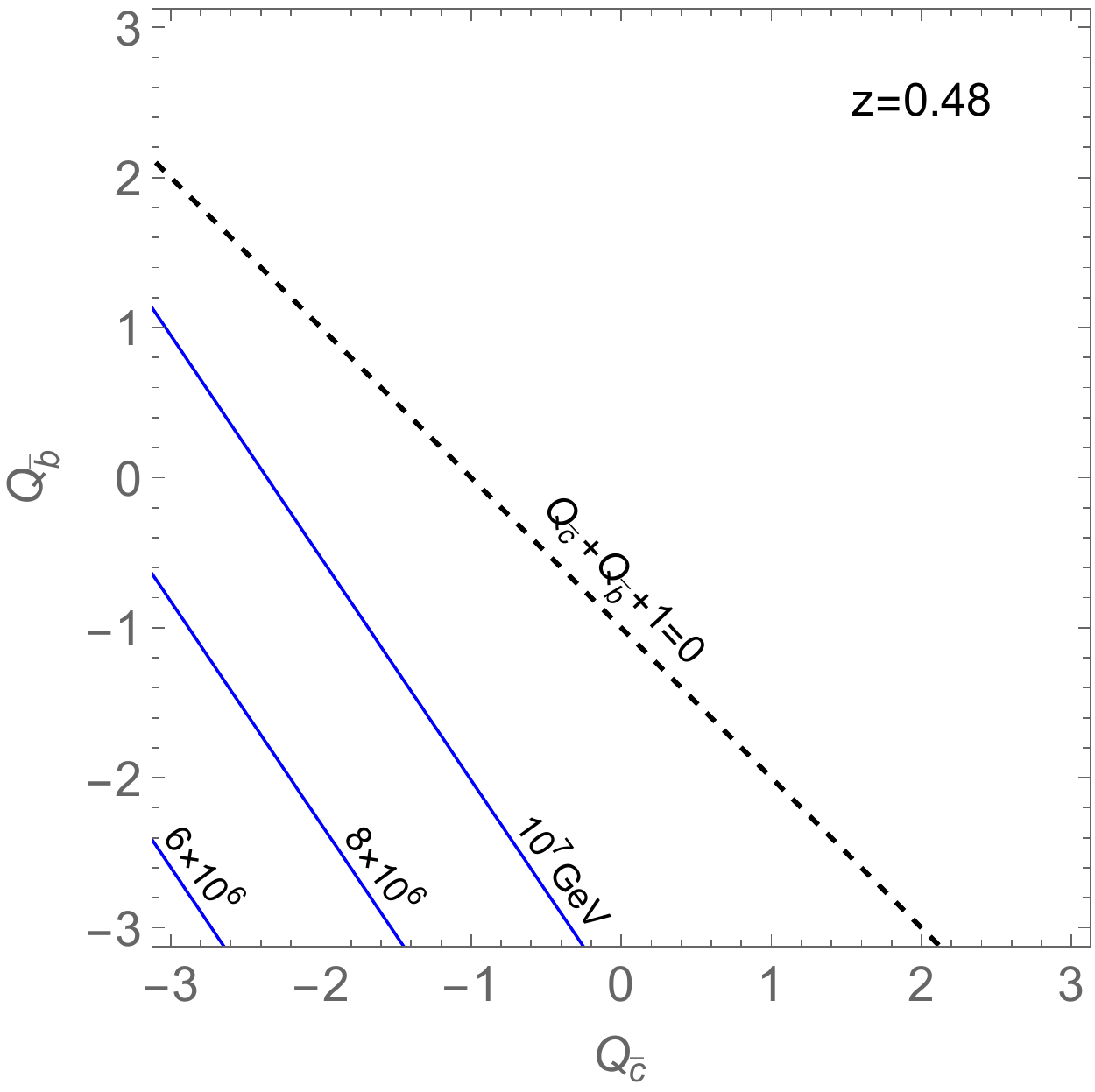}
	\includegraphics[width=0.49\linewidth]{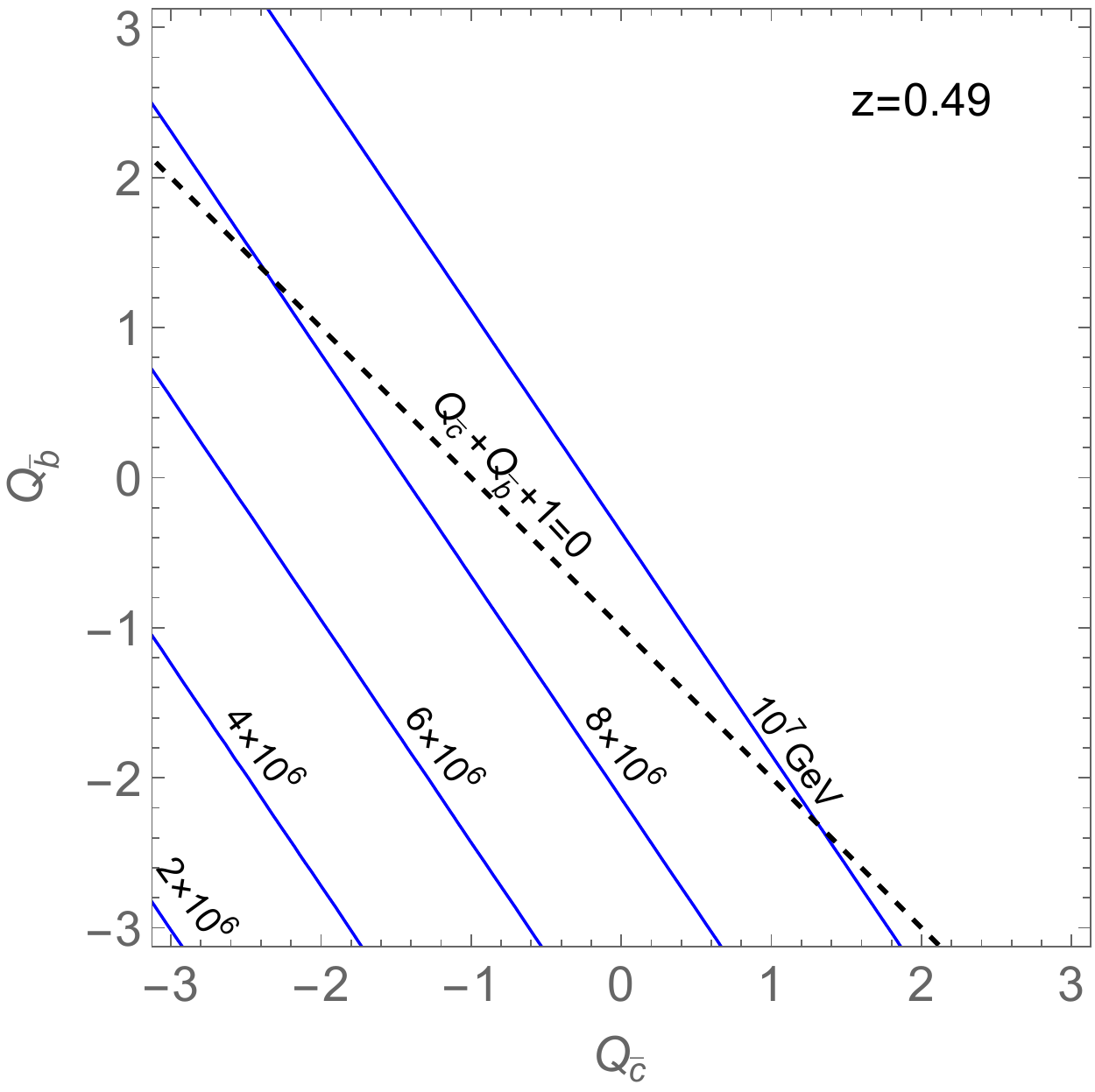}
 	\includegraphics[width=0.49\linewidth]{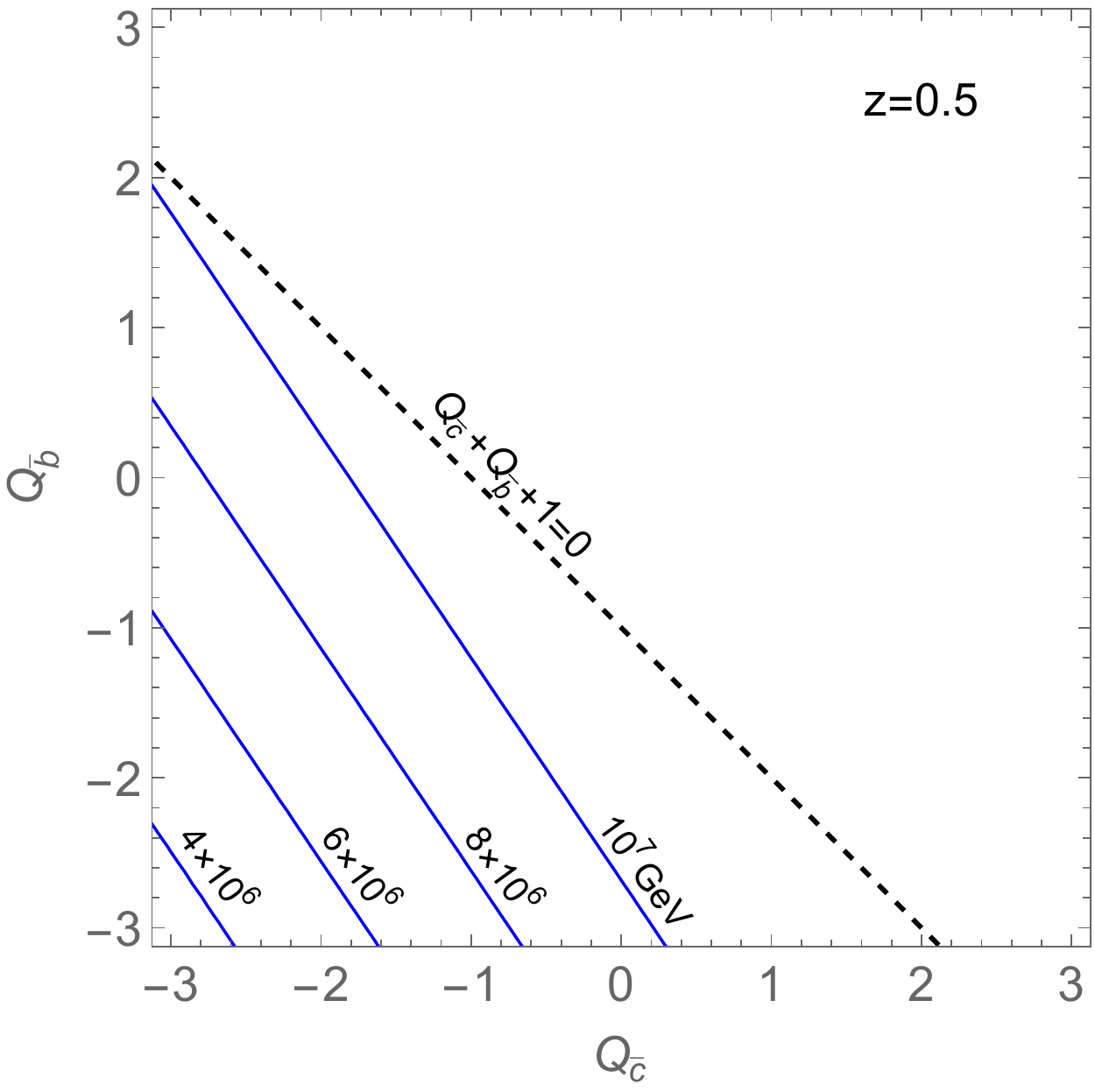}
   	\includegraphics[width=0.49\linewidth]{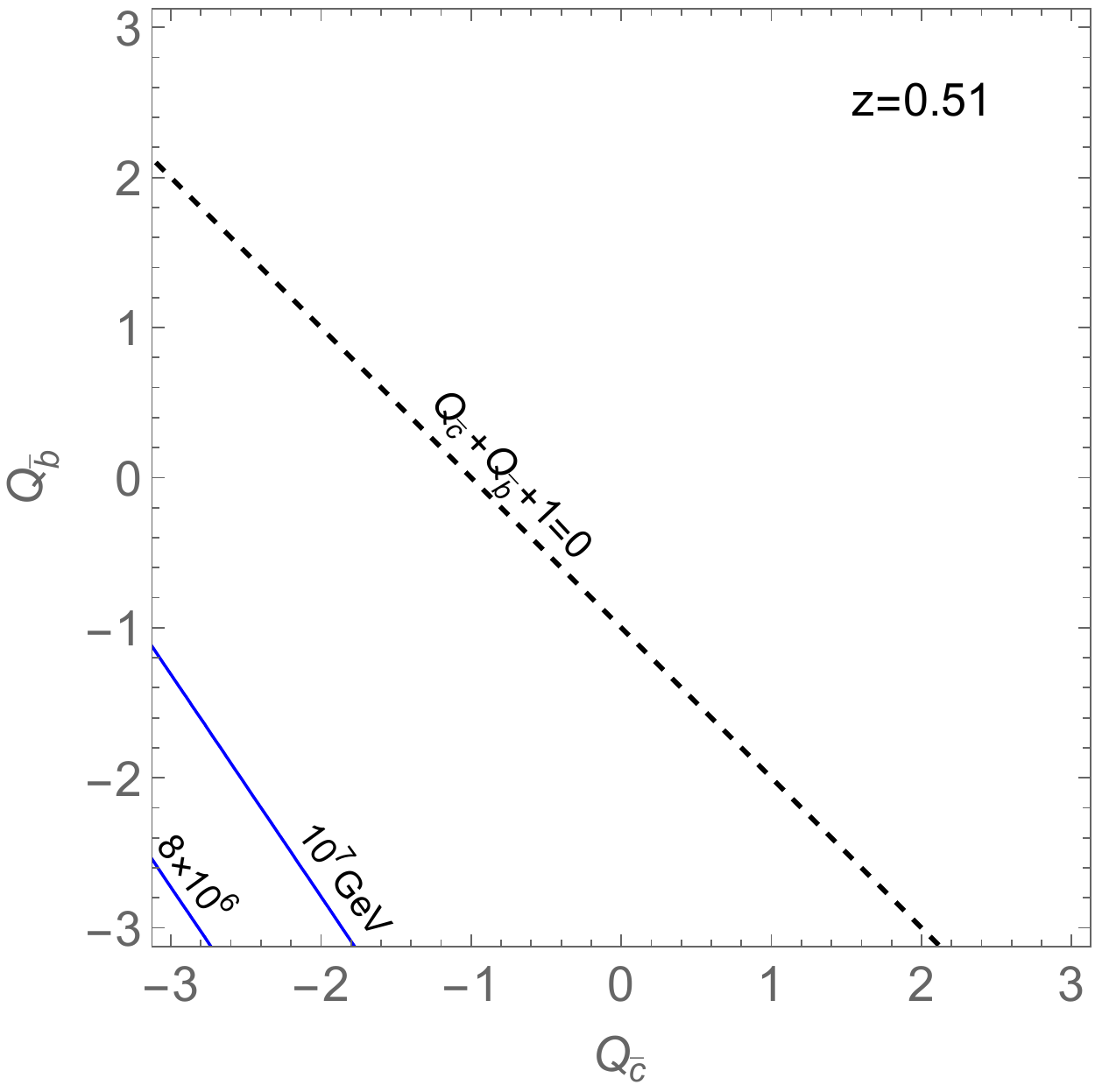}
	\caption{The lower bound on the decay constant $f_a$ from the cooling of neutron stars for $Q_{\bar{s}}=1$
 and $G=0$. The dashed lines show the case where the QCD anomaly of the PQ symmetry is determined by the SM quarks.}
	\label{fig:nucleon}	
\end{figure}

\begin{figure}[!t]
\centering
 \includegraphics[width=0.49\linewidth]{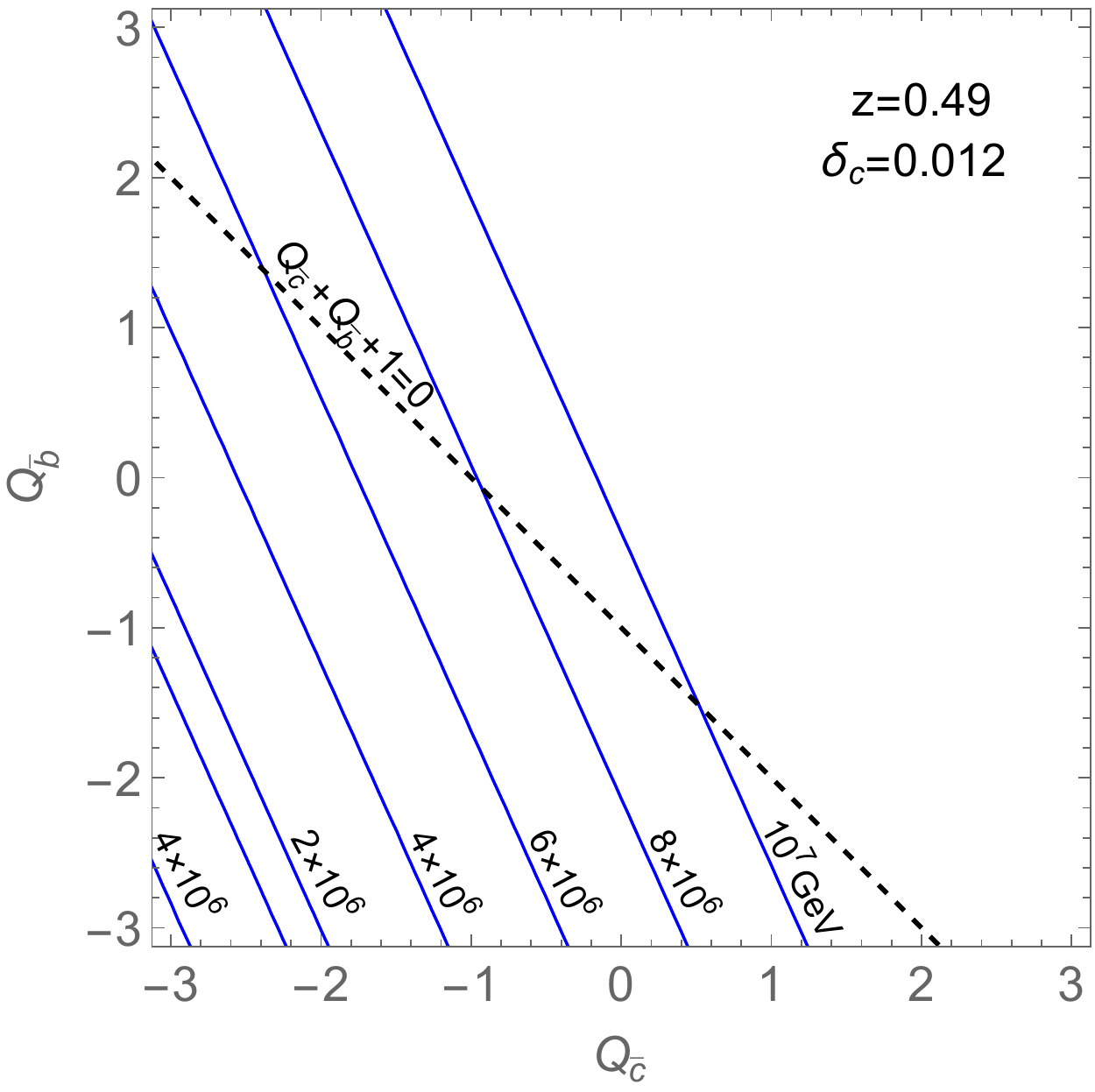}
\includegraphics[width=0.49\linewidth]{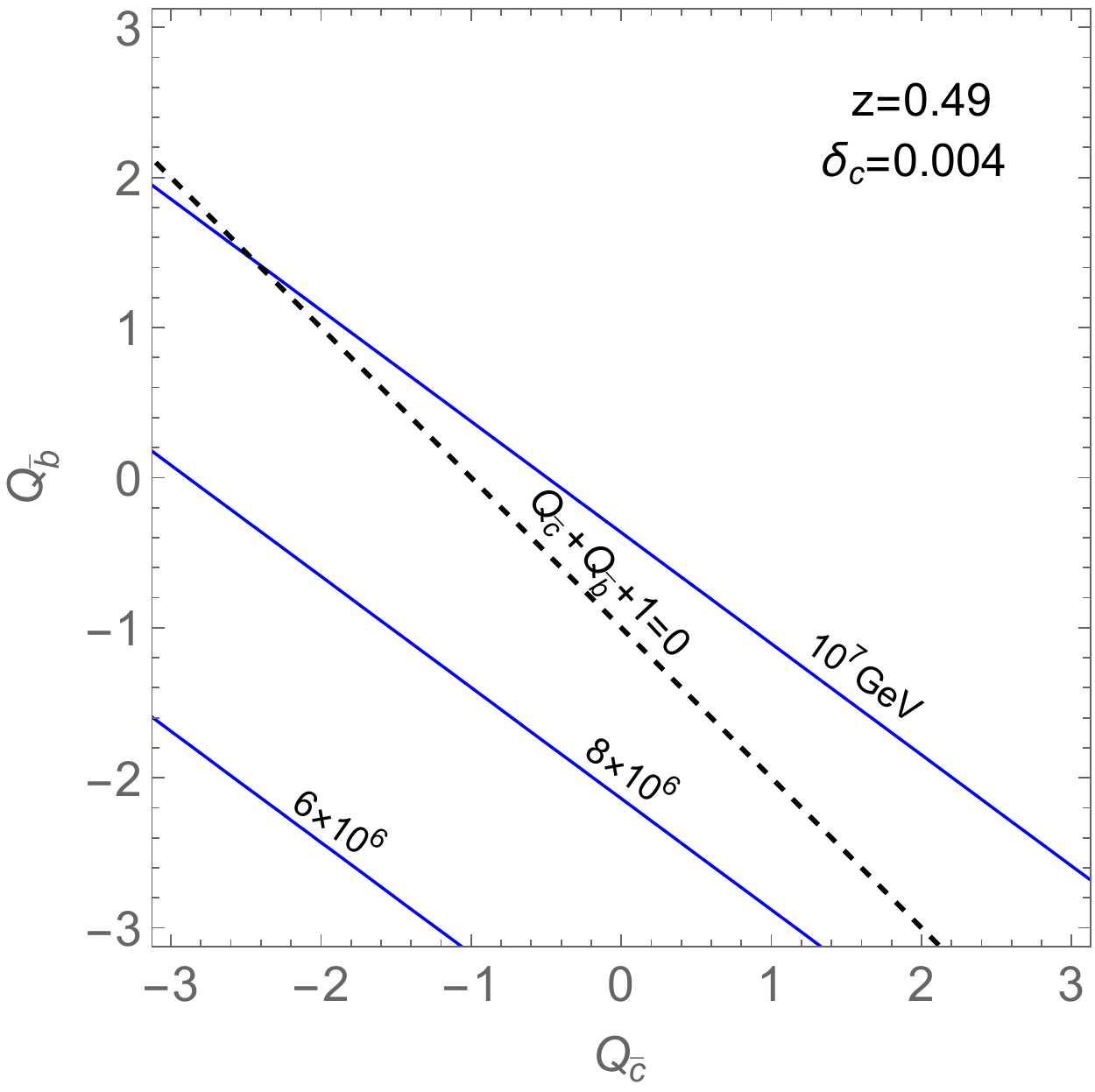}
	\caption{Same as Fig.~\ref{fig:nucleon}, but with different $\delta_c$.}
	\label{fig:nucleon2}	
\end{figure}  

In the minimal model, $Q_{\bar{s}} \neq Q_{\bar{d}}$.  
As we will see in Sec.~\ref{sec:flavor}, the flavor-violating axion-down-strange coupling is generically too large unless special flavor structure is imposed. We thus also consider models with $Q_{\bar{s}}= Q_{\bar{d}} =1$ and allow any SM quarks except for the top quark to have non-zero PQ charges, as shown in Table~\ref{tab:charge_nonmin}.
If only SM quarks contribute to the QCD anomaly, $Q_{\bar{c}}+ Q_{\bar{b}}=-1$ 
is required. If there are extra colored particles that contribute to the QCD anomaly, as in some of the UV completion introduced in Sec.~\ref{sec:UV}, they can take different values. In these models, it is possible that some of the SM fermions are not PQ charge eigenstates and their charges are only effective ones that are not quantized; see Sec.~\ref{sec:UV}.

\begin{table}[!t]
    \centering
    \begin{tabular}{|c|c|c|c|c|c|c|} \hline
      & $\bar{u}$ & $\bar{c}$ & $\bar{t}$ & $\bar{d}$ & $\bar{s}$ & $\bar{b}$    \\ \hline
      $Q_f$ & $2$ & $Q_{\bar{c}}$ & $0$ & $1$ & $1$ & $Q_{\bar{b}}$ \\ \hline
    \end{tabular}
    \caption{The PQ charges $Q_f$ of the SM quarks that avoid the flavor bound without relying on special flavor structure. If only SM quarks contribute to the QCD anomaly, 
    $Q_{\bar{b}}+Q_{\bar{c}} +1 =0$ 
    is required for the axion-nucleon coupling to be suppressed.}
    \label{tab:charge_nonmin}
\end{table}

In Fig.~\ref{fig:nucleon}, we show the lower bound on $f_a$ from the cooling of neutron stars as a function of $(Q_{\bar{c}},Q_{\bar{b}})$ for $Q_{\bar{s}}=1$ and $G=0$. For reasonable choice of $Q_{\bar{c}}$ and $Q_{\bar{b}}$, $f_a$ may be below $10^7$~GeV, and as small as $2\times 10^6$~GeV for $z\simeq 0.49$.

In Fig.~\ref{fig:nucleon}, the central values for $\delta_{s,c,b}$ and $G=0$ are assumed. Using different values do not change how low $f_a$ can be for given $z$ once scanned over $(Q_{\bar{c}},Q_{\bar{b}})$, but the preferred set of $(Q_{\bar{c}},Q_{\bar{b}})$ changes. In Fig.~\ref{fig:nucleon2}, we show the same plots as Fig.~\ref{fig:nucleon} but with different $\delta_c$ that has the largest fractional error.
The coefficient $\delta_s$ also has a large absolute error. Changing it within the error shifts the contours of the lower bound on $f_a$ by $\pm 0.008/ (0.008(4))$ to the $Q_{\bar{c}}$ direction. The effect of different $G$ can be also estimated in the similar manner. Numerically, the contribution of $G$ to $\delta$ is $0.001\times G$, so the effect of its uncertainty is subdominant.

The constraint from the cooling of neutron stars may be relaxed if the heating of neutron stars by the decay of magnetic fields in them is significant~\cite{,Buschmann:2021juv}.
The bound from SN1987A, however, gives a similar constraint~\cite{Carenza:2019pxu}.

\subsection{Lepton couplings}

The axion-lepton couplings are also given by Eqs.~\eqref{eq:quarkcoup} and \eqref{eq:coupling_dirac}.
To suppress the coupling with the electron,  $Q_{\bar{e}}$ should be zero. 
We also assume that $Q_{\bar{\mu}}=0$; otherwise the lower bound on $f_a$ from SN1987A is $\mathcal{O}(10^8)$ GeV~\cite{Bollig:2020xdr}. This also suppresses the muon-electron flavor-violating axion coupling.
So the only possible PQ-charged lepton is the right-handed tau. In the minimal model, $Q_{\bar{\tau}}=0$.

\subsection{Photon coupling}

The axion-photon coupling is given by
\begin{align}
- \frac{g_{a\gamma\gamma}}{8} a \epsilon^{\mu\nu\rho\sigma} F_{\mu\nu}F_{\rho\sigma},~~~&
g_{a\gamma\gamma}= \frac{\alpha}{2\pi f_a}C_\gamma,\nonumber \\
&C_\gamma = \left(\frac{E}{N} - C_{\gamma}^{\rm QCD}\right),~~C_{\gamma}^{\rm QCD} = \frac{2}{3}\frac{4+z+w}{1+z} + 0.06(2), 
\end{align}
where $E/N$ is the electromagnetic anomaly coefficient of the PQ symmetry relative to the QCD anomaly. Here we used the estimation in~\cite{Lu:2020rhp} with keeping $z$ and $w$ unfixed. For the central values of these parameters, $C_\gamma^{\rm QCD} \simeq  2.07(4)$.   The observations of the stellar population in globular clusters gives~\cite{Ayala:2014pea}
\begin{align}
f_a > 1.8 \times 10^7 ~{\rm GeV} \times \left| C_\gamma \right| \,.
\end{align}
For $E/N\neq 2$, $f_a$ should be above $\mathcal{O}(10^7)$ GeV.

To have $f_a = \mathcal{O}(10^6)$ GeV, as required for successful minimal axiogenesis discussed in Sec.~\ref{sec:axiogenesis}, it is necessary to suppress the axion-photon coupling. This can be naturally achieved by $E/N=2$~\cite{Kaplan:1985dv}, which leads to $|C_\gamma| = \mathcal{O}(0.1)$ and $f_a \gtrsim  10^6$ GeV.%
\footnote{The result for the axion-photon coupling neglecting the effects of the strange quark, i.e., with $w=0$, found in~\cite{GrillidiCortona:2015jxo}, $C_\gamma \simeq E/N -1.92(4)$, is significantly different from that in \cite{Lu:2020rhp}. However, for $E/N=2$, a very similar lower bound on $f_a$ is obtained.}
Interestingly, in the minimal model, if there is no electromagnatic anomaly of the PQ symmetry beyond that given by $\bar{u}$ and $\bar{d}$, $E/N=2$ is satisfied.
More generically, if the QCD and electromagnetic anomaly of the PQ symmetry is only given by the SM fermions, the PQ charges of them are constrained, as shown in Table~\ref{tab:charge_min}.

\begin{table}[!t]
    \centering
    \begin{tabular}{|c|c|c|c|c|c|c|c|c|c|}
    \hline
      & $\bar{u}$ & $\bar{c}$  & $\bar{t}$ &  $\bar{d}$ & $\bar{s}$  & $\bar{b}$ &  $\bar{e}$ & $\bar{\mu}$  & $\bar{\tau}$    \\ \hline
      $Q_f$ & $2$  & $Q_{\bar{c}}$ & $0$ & $1$ & $Q_{\bar{s}}$ & $-(Q_{\bar{c}}+Q_{\bar{s}})$ & $0$  & $0$ & $-Q_{\bar{c}}$ \\ \hline
    \end{tabular}
    \caption{The PQ charges $Q_f$ of the SM fermions that lead to naturally suppressed axion-nucleon and axion-photon couplings. Here it is assumed that the QCD and electromagnetic anomaly of the PQ symmetry solely comes from the SM fermions.}
    \label{tab:charge_min}
\end{table}

\subsection{Flavor violation}
\label{sec:flavor}

The setup generically leads to flavor violation. Let us take the first two generations of the up-type quarks. The Yukawa interactions are given by
\begin{align}
  H
  \begin{pmatrix}
  q_1 & q_2
  \end{pmatrix}
  \begin{pmatrix}
  y_{11}e^{-2i\theta} & y_{12}e^{-iQ_{\bar{c}}\theta} \\
  y_{21}e^{-2i\theta} & y_{22}e^{-iQ_{\bar{c}}\theta} \\  
  \end{pmatrix}
  \begin{pmatrix}
  \bar{u} \\ \bar{c}
  \end{pmatrix} + {\rm h.c.},
\end{align}
where $q_1$ and $q_2$ are the first- and second-generation quark doublets, respectively.
We may remove the axion from the mass matrix by the rotation of $\bar{u}$ and $\bar{c}$,
\begin{align}
-\partial_\mu \theta 
\begin{pmatrix}
\bar{u}^\dag & \bar{c}^\dag
\end{pmatrix}
\begin{pmatrix}
2 & \\
 & Q_{\bar{c}}
\end{pmatrix}
\bar{\sigma}^\mu
\begin{pmatrix}
\bar{u} \\
\bar{c}
\end{pmatrix}
+
  H
  \begin{pmatrix}
  q_1 & q_2
  \end{pmatrix}
  \begin{pmatrix}
  y_{11} & y_{12}\\
  y_{21} & y_{22} \\  
  \end{pmatrix}
  \begin{pmatrix}
  \bar{u} \\ \bar{c}
  \end{pmatrix} + {\rm h.c.}
\end{align}
The quark mass matrix is diagonalized by the rotation of $(u,c)$ and $(\bar{u},\bar{c})$. Unless the PQ charge of $\bar{u}$ is equal to that of $\bar{c}$, i.e., $Q_{\bar{c}}=2$, the rotation introduces an axion-$\bar{u}$-$\bar{c}$ coupling.
The flavor violation is suppressed in the following structure that is consistent with generic flavor symmetry that explains $y_u \ll y_c$~\cite{Co:2022aav},
\begin{align}
y_{11}\simeq y_u,~y_{22}\simeq y_c,~~y_{21} = \epsilon_u y_u,~~y_{12} = \epsilon_c y_c,~~\epsilon_{u,c} \sim \theta_{12} \sim 0.1,
\end{align}
where $\theta_{12}$ is the CKM mixing between the first and second generations. The rotation angle $\theta_{\bar{u} \bar{c}}$ between $\bar{u}$ and $\bar{c}$ to diagonalize the mass matrix is
\begin{align}
\theta_{\bar{u} \bar{c}} = (\epsilon_u + \epsilon_c ) \frac{y_u}{y_c}= \mathcal{O}(10^{-4}).
\end{align}
With this suppression, the lower bound on the axion decay constant from $D$-meson decay is $\mathcal{O}(10^4)$ GeV. Here we used the constraint derived in~\cite{MartinCamalich:2020dfe} based on~\cite{CLEO:2008ffk}.
Note that the bound is only $\mathcal{O}(10^5)$ GeV even if $\epsilon_{u,c} = \mathcal{O}(1)$, so a flavor symmetry that explains the Cabibbo angle is not necessary as far as the flavor violation is concerned.
Assuming similar suppression of flavor-violating axion-bottom couplings, the lower bound on $f_a$ from $B$-meson decay~\cite{MartinCamalich:2020dfe,BaBar:2013npw} for $Q_{\bar{b}} \neq Q_{\bar{s}}$ is also $\mathcal{O}(10^5)$ GeV.

If $Q_{\bar{d}}\neq Q_{\bar{s}}$, axion-$\bar{d}$-$\bar{s}$ coupling is introduced. Applying the same analysis as the $\bar{u}$-$\bar{c}$ case, the expected suppression is $\theta_{12} y_d/y_s = \mathcal{O}(10^{-2})$. The lower bound on $f_a$ from kaon decay~\cite{MartinCamalich:2020dfe,E949:2007xyy} would be then $10^{10}$ GeV, which is much stronger than the astrophysical bounds. This bound is avoided by $Q_{\bar{s}} = Q_{\bar{d}}$. In the minimal model, however, $Q_{\bar{s}} = 0 \neq Q_{\bar{d}}$.
The flavor violation can be suppressed by imposing a particular flavor symmetry and its breaking. For example, we may impose $U(1)_d\times U(1)_s$ symmetry with charges $q_1(1,0)$, $q_2(0,1)$, $\bar{d}(-1,0)$, and $\bar{s}(0,-1)$  
so that the coupling of the axion with the down and strange quarks becomes flavor diagonal. The CKM mixing comes from the up-type Yukawa that explicitly breaks this symmetry. Using the up-type Yukawa as a spurion, one can show that the axion-down-strange coupling is suppressed by $\theta_{12} y_c^2 y_d/y_s \sim 10^{-7}$, and the lower bound on $f_a$ is $\mathcal{O}(10^5)$ GeV.

If $Q_{\bar{\tau}}\neq 0$, flavor-violating axion-tau-muon and electron couplings are introduced. Assuming the similar suppression as the quark sector, the strongest constraint comes from the axion-tau-muon coupling that is suppressed only by $y_\mu/y_\tau = \mathcal{O}(0.1)$; the 2-3 MNS mixing is $\mathcal{O}(1)$. Using the bound derived in~\cite{Calibbi:2020jvd} based on~\cite{ARGUS:1995bjh}, the lower bound on $f_a$ from tau decay is $\mathcal{O}(10^5)$ GeV.
Interestingly, the strongest constraint on the axion-tau-muon coupling comes from cosmology due to hot axions produced in flavor-violating tau decays in the early Universe. The Planck constraint on dark radiation~\cite{Planck:2018vyg} leads to the lower bound on $f_a$ of $\mathcal{O}(10^6)$~GeV~\cite{DEramo:2021usm}. 

If the left-handed quarks have non-universal PQ charges, the flavor-violating axion couplings are suppressed only by the CKM angle and the lower bound on $f_a$ becomes stronger. For example, if $Q_{q_1}\neq Q_{q_2}$, the lower bound on $f_a$ from kaon decay is $\mathcal{O}(10^{11})$ GeV.
Even with flavor symmetry that eliminates axion-strange-down coupling, axion-charm-up coupling is unavoidable, and the lower bound on $f_a$ is $\mathcal{O}(10^7)$ GeV.
If $Q_{q_3}\neq Q_{q_2}$, the lower bound on $f_a$ from $B$-meson decay is $\mathcal{O}(10^6)$ GeV. Although this is not as stringent as the astrophysical bound, the constraint is much stronger than the case with $Q_{\bar{b}} \neq Q_{\bar{s}}$ and $Q_{q_3}=Q_{q_2}$.

\section{UV completions}
\label{sec:UV}

In this section, we present UV completions of the axion-dependent Yukawa couplings. They may be understood as higher dimensional couplings between the SM fields and the PQ breaking field $P$, which can be generated by the exchange of heavy fermions or scalars. We denote the vacuum expectation value of $P$ as $v_P$ and the phase direction of $P$ as $\theta_P$.

\subsection{Extra fermions}
\label{sec:UVfermions}

Let us first discuss the axion-dependent up Yukawa. It may be UV completed by introducing vector-like fermions that have the same gauge charge as the right-handed up quark, $U_{-1}$ and $\bar{U}_0$, where the subscripts denote the PQ charges. The quark that is the dominant component of the right-handed up quark is denoted as $\bar{u}_2$. The Yukawa couplings consistent with the PQ symmetry are
\begin{align}
\label{eq:up1}
y H q \bar{U}_0 + \left( \lambda_0 P \bar{U}_0 + \lambda_2P^\dag \bar{u}_2 \right) U_{-1} + {\rm h.c.}, 
\end{align}
where we assume that the PQ charge of $P$ is $1$.
Assuming $\lambda_0 \gg \lambda_2$, we may integrate out the heavy fermions $\bar{U}_0$ and $U_{-1}$ to obtain an effective coupling
\begin{align}
-  y_u e^{-2i \theta_P} H q \bar{u},~~y_u = \frac{y \lambda_2}{\lambda_0},
\end{align}
where $\bar{u}$ is the right-handed up quark that is dominantly $\bar{u}_2$.
In this setup, the heavy fermions $\bar{U}_0$ and $U_{-1}$ contribute to the QCD anomaly of the PQ symmetry. Also, the small up Yukawa coupling is not explained by the ratio between the PQ symmetry and some other higher mass scale, but rather by small couplings $y \lambda_2$, which can be understood by some flavor symmetry.

One may consider a more Froggatt-Nielsen-like PQ model by introducing $U_{-1}$, $\bar{U}_1$, $\bar{U}_0$, and $U_0$. If their Dirac masses $M$ are larger than the PQ symmetry-breaking scale, the up Yukawa coupling is suppressed by $v_P^2/M^2$. In this case, the extra fermions do not contribute to the QCD anomaly.

The extra fermions may have the same gauge charges as doublet quarks. Let us introduce $Q_{-2}$ and $\bar{Q}_1$ and the couplings
\begin{align}
\label{eq:up2}
y H Q_{-2} \bar{u} + \left( \lambda_{-2} P Q_{-2} + \lambda_0 P^\dag q \right) \bar{Q}_{1} + {\rm h.c.} 
\end{align}
After integrating out $Q_{-2}$ and $\bar{Q}_1$ assuming $\lambda_{-2} \gg \lambda_0$, we obtain
\begin{align}
-  y_u e^{-2i \theta_P} H q \bar{u},~~y_u = \frac{y \lambda_0}{\lambda_{-2}}.
\end{align}
In this setup, the extra fermions contribute to both the QCD and weak anomaly of the PQ symmetry.
It is also possible to make the model more Froggatt-Nielsen like by further introducing $Q_{-1}$ and $\bar{Q}_2$.

We next discuss the down Yukawa. We may introduce $D_0$ and $\bar{D}_0$ that has the same gauge charge as the right-handed down quark and the couplings and the mass term
\begin{align}
\label{eq:down1}
y H^\dag q \bar{D}_0 + \lambda P^\dag D_0 \bar{d}_1 + M D_0 \bar{D}_0 + {\rm h.c.}
\end{align}
Assuming $M\gg \lambda v_P$, we may integrate out $D_0$ and $\bar{D}_0$ to obtain
\begin{align}
- y_d e^{-i \theta_P} H^\dag q \bar{d},~~ y_d = \frac{y\lambda v_P }{M}.
\end{align}
In this setup, the small down Yukawa may be understood by  small $v_P/M$. Also, the extra fermions do not contribute to the QCD anomaly.
We may instead introduce $D_{-1}$ and $\bar{D}_0$ and the couplings and the mass
\begin{align}
y H^\dag q \bar{D}_0 + \lambda P \bar{D}_{0} D_{-1} + M D_{-1} \bar{d}_1 + {\rm h.c.}
\end{align}
Assuming $\lambda v_P \gg M$, we may integrate out $D_{-1}$ and $\bar{D}_0$ to obtain
\begin{align}
- y_d e^{-i \theta_P} H^\dag q \bar{d},~~ y_d = \frac{y M }{\lambda v_P}.
\end{align}
In this setup, the small down Yukawa may be understood by  small $M/v_P$. The extra fermions contribute to the QCD anomaly.
It is straightforward to construct UV completion of the down Yukawa by $Q_{-1}$ and $\bar{Q}_1$, and that by $Q_{-1}$ and $\bar{Q}_0$.

Similar UV completion can be straightforwardly constructed for other Yukawa couplings.

We note that the UV completion by extra fermions can produce axion-SM fermions couplings that are not quantized if mass parameters of the theory are not hierarchically different from each other. For example, let us consider the UV completion in Eq.~\eqref{eq:down1} beyond the approximation $M \gg \lambda \vev{P}$. We replace $P$ with $v_P e^{i \theta_P}$ and remove $\theta_P$ from the mass term by the rotation of $\bar{d}_1$. 
A linear combination of $\bar{D}_0$ and $\bar{d}_1$, which we denote as $\bar{D}$, obtains a large Dirac mass $\sqrt{M^2 + \lambda^2 v_P^2}$ paired with $D_0$. The relation between $(\bar{d},\bar{D})$ and $(\bar{d}_1,\bar{D}_0)$ is given by
\begin{align}
\begin{pmatrix}
\bar{d}_1 \\ \bar{D}_0
\end{pmatrix} =
\begin{pmatrix}
    {\rm cos}\alpha & {\rm sin} \alpha \\
    -{\rm sin} \alpha & {\rm cos} \alpha
\end{pmatrix}
\begin{pmatrix}
    \bar{d} \\ \bar{D}
\end{pmatrix},~~
{\rm tan}\alpha = \frac{\lambda v_P}{M}.
\end{align}
The coupling of $\theta_P$ with $\bar{d}$ originates from that with $\bar{d}_1$ and is given by
\begin{align}
-\partial_\mu \theta_P \bar{d}^\dag \bar{\sigma}^\mu \bar{d} \times {\rm cos}^2 \alpha,
\end{align}
which is not quantized. This is because the SM right-handed down quark is a linear combination of fermions with different PQ charges. If $M\gg \lambda v_P$ or $M\ll \lambda v_P$, for which the SM right-handed down quark is almost a PQ charge eigenstate, the axion-down coupling is almost quantized. 
Generically, if a SM fermion $f$ is a linear combination of $f_i$ with PQ charges $Q_{f_i}$ with coefficients $c_i$, the axion-fermion coupling is determined by an effective PQ charge
\begin{align}
 Q_{{\rm eff},f} =\sum_i|c_i|^2Q_{f_i}.
\end{align}
Unless $f$ is almost a PQ charge eigenstate, the axion-fermion coupling is no longer quantized.

The suppressed axion-nucleon coupling requires $C_u\simeq 2/3$ and $C_d \simeq 1/3$. To achieve this without fine-tuning, it is crucial that the up and down quarks are nearly PQ charge eigenstates. To suppress the flavor-violating axion-down-strange coupling, electron coupling, and muon coupling, the strange quark, electron, and muon should also be approximately PQ charge eigenstates. The charm, bottom, and top quark and the tau do not have to have quantized coupling with the axion, so they may be a mixture of different PQ charge eigenstates. Irrational $(Q_{\bar{c}},Q_{\bar{b}})$ in Figs.~\ref{fig:nucleon} and \ref{fig:nucleon2} may be understood in this way. This generically requires coincidence of masses, but in the UV completion in Eqs.~\eqref{eq:up1} and \eqref{eq:up2}, the masses are given by the PQ breaking field and the coincidence is required for dimensionless constants rather than for energy scales.

\subsection{Extra scalars}

Let us discuss the up Yukawa. We introduce a scalar field that has the same gauge charge as the SM Higgs and has a PQ charge of $-2$, $H_{-2}$. The interactions and masses consistent with the PQ symmetry is
\begin{align}
\left(y H_{-2}q \bar{u} + \lambda P^2H_{-2} H^\dag + {\rm h.c.}\right) - M^2|H_{-2}|^2.
\end{align}
Assuming $M^2 \gg \lambda v_P^2$, we may integrate out $H_{-2}$ to obtain
\begin{align}
  y_u e^{-2i \theta_P} H q \bar{u},~~y_u = \frac{y \lambda v_P^2}{M^2}.
\end{align}

Other Yukawa couplings can be obtained in a similar manner by introducing scalar fields that have the same gauge charge as the SM Higgs and appropriate PQ charges. In this setup, the gauge anomaly of the PQ symmetry solely comes from the SM fermions.
The minimal model can be UV-completed by three Higgses; $H$, $H_{-2}$, and $H_1$.

If the SM Higgs is not nearly a zero-PQ charge eigenstate, in the low energy EFT after integrating out heavy Higgses, the axion couples to the Higgs current. This may be removed by the hypercharge rotation proportional to the axion field without gauge transformation on gauge fields, but axion-fermion current couplings proportional to the hypercharge are induced. This gives $C_u/C_d \neq 2$ and $C_e \neq 0$, so the axion would be no longer astrophobic. It is necessary that the SM Higgs is nearly a zero-PQ charge eigenstate. This means that the $\mathcal{O}(1)$ top yukawa should be PQ neutral.

\section{Stellar cooling anomalies}
\label{sec:cooling}

Let us also discuss stellar-cooling anomalies that have been observed in several stellar environments and can be explained if the axion couples to electrons~\cite{Giannotti:2017hny}. The axion-electron coupling required to explain these anomalies is $C_e\simeq2\times10^{-3}f_a/(10^7\,\GeV)$, with the SM disfavored by more than~$3\sigma$~\cite{DiLuzio:2021ysg}. Such a value prefers models with either tree-level axion-electron coupling and/or small $f_a$. This cannot be achieved in the minimal KSVZ model while in the minimal DFSZ model explanation of these anomalies is in tension with stringent astrophysical constraints on axion-nucleon couplings or the perturbativity of Yukawa couplings~\cite{Giannotti:2017hny}.  On the other hand, the stellar-cooling anomalies can be perfectly explained in nucleophobic models with tree-level axion-electron couplings~\cite{DiLuzio:2017ogq,Bjorkeroth:2019jtx,Saikawa:2019lng,Badziak:2021apn,DiLuzio:2021ysg}. 

While the axion-photon coupling is not necessary to explain these anomalies, 
the best-fit value of the axion-photon coupling is  $|C_\gamma|\simeq f_a/(6\times10^7\,\GeV)$~\cite{DiLuzio:2021ysg}. For typical values of $E/N$, $f_a$ is preferred to be between $10^7$ and $10^8$~GeV while for $E/N=2$, $f_a\simeq \mathcal{O}(10^6)$~GeV is preferred. Such small $f_a$ is viable only in models with suppressed axion-nucleon couplings. In this range of $f_a$ the best-fit is obtained for $C_e\simeq \mathcal{O}(10^{-2})$ if $E/N\neq2$ or $C_e\simeq \mathcal{O}(10^{-3})$ if $E/N=2$.

In our setup, a small tree-level electron coupling $C_e^{\rm tree}$ arises if the electron or the SM Higgs contains a small fraction of a state with a non-zero PQ charge, as discussed in Sec.~\ref{sec:UV}.
We note that $C_e^{\rm tree}\neq0$ generically leads to $\mu\to e a$ decay and the lower bound on $f_a$ from this decay, found in \cite{Calibbi:2020jvd} based on \cite{Jodidio:1986mz,TWIST:2014ymv}, is $\mathcal{O}(C_e^{\rm tree} 10^7\,(10^9)\,\GeV)$  with (without) suppression of the flavor violation by $y_e/y_\mu$. Thus, when $C_e^{\rm tree}$ originates from the effective PQ charge of the right-handed electron the flavor violation does not lead to any relevant constraint for values of $C_e^{\rm tree}$ and $f_a$ that explain stellar-cooling anomalies. On the other hand, if the flavor-violation comes from the effective PQ charge of the left-handed electron, the value of $f_a$ that explains the stellar-cooling anomalies is still comparable with the lower bound on $f_a$ from $\mu\to e a$. The latter scenario is expected to be tested with future muon beam experiments~\cite{Calibbi:2020jvd} such as MEG II~\cite{MEGII:2018kmf} and Mu3e~\cite{Berger:2014vba,Blondel:2013ia}.

It is also possible to generate an axion-electron coupling by quantum corrections~\cite{Chang:1993gm}. In the minimal model, the RG correction between $f_a$ and the QCD scale $\Lambda_{\rm QCD}$ is proportional to the up or down yukawa coupling and is negligible. The quantum correction between the QCD scale and $m_e$ is given by
\begin{align}
    C_e \simeq  \frac{3\alpha^2}{4 \pi^2}C_\gamma\log\left(\frac{\Lambda_{\rm QCD}}{m_e}\right) \simeq  2 \times 10^{-5}C_\gamma,
\end{align}
which is too small to explain the cooling anomaly.

In non-minimal models, the quantum correction can be larger. For example, an axion-$W$ boson coupling that can arise from heavy $SU(2)$- and PQ-charged fermions gives
\begin{align}
C_e \simeq  \frac{9}{2}\frac{\alpha_2^2}{16\pi^2}c_W \log\left(\frac{f_a}{m_W}\right)  \simeq  3\times 10^{-4} c_W \frac{\log(f_a/m_W)}{10}, 
\end{align}
where $c_W$ is the weak anomaly of the PQ symmetry relative to the QCD anomaly and we used the RGE in~\cite{DiLuzio:2022tyc}. 
Here we assumed a RGE correction from $f_a$ to maximize the correction; generically the correction starts from the masses of heavy fermions responsible for the weak anomaly, which may be smaller than $f_a$ if the coupling of them with the PQ breaking field is small.
Comparing this with the result in~\cite{DiLuzio:2021ysg}, one can see that this electron coupling can explain the cooling anomaly within $2 \sigma$ when $E/N\neq 2$ and $f_a = {\cal O}(10^{7})$ GeV, and $1 \sigma$ when $E/N=2$ and $f_a = {\cal O}(10^{6})$ GeV. Such small values of $f_a$ are viable only for astrophobic axions.

An axion-gluon coupling arising from heavy PQ-charged colored fermions also generates $C_e$ through a two-loop correction involving the top Yukawa coupling to the axion-Higgs coupling. It is of the similar size as the one from the axion-$W$ boson coupling with $c_W\sim1$~\cite{Bauer:2020jbp} and can explain the cooling anomalies.

The quantum corrections involving the hypercharge gauge interaction tend to be small. The RGE correction between $f_a$ and $m_W$ is given by
\begin{align}
C_e \simeq& \frac{15}{2}\frac{\alpha_1^2}{16\pi^2}\left(c_Y + \frac{8}{3}(c_{\bar{u}} + c_{\bar{c}}) + \frac{2}{3}(c_{\bar{d}} + c_{\bar{s}} + c_{\bar{b}}) \right)\log\left(\frac{f_a}{m_W}\right) \nonumber \\
 \simeq & 5\times 10^{-5} \left(c_Y + \frac{8}{3}(c_{\bar{u}} + c_{\bar{c}}) + \frac{2}{3}(c_{\bar{d}} + c_{\bar{s}} + c_{\bar{b}}) \right) \frac{\log(f_a/m_W)}{10},
\end{align}
where $c_Y$ is the hypercharge anomaly of the PQ symmetry relative to the QCD anomaly.
For $E/N\neq 2$, this electron coupling cannot explain the cooling anomaly. For $E/N=2$, the anomaly can be explained within $2\sigma$.

\section{Minimal axiogenesis}
\label{sec:axiogenesis}

In this section, we discuss the compatibility of the astrophobic axion model with the baryogenesis scenario from the rotation of the axion field and the electroweak sphaleron process, dubbed as minimal axiogenesis~\cite{Co:2019wyp}.

If the radial direction of the PQ-breaking field $P$ is flat, as naturally occurs in supersymmetric theories, the radial direction may take on a large field value in the early universe. Then higher order terms in the potential of $P$ becomes important. We assume that some of them explicitly violate the PQ symmetry, so that rotation of $P$ in field space is initiated by the potential gradient to the angular direction, as in the Affleck-Dine mechanism~\cite{Affleck:1984fy}. It is also possible to first initiate the rotation of other scalar fields, such as squarks and sleptons, and transfer the angular momentum of them to $P$~\cite{Domcke:2022wpb}. In this case, the potential of $P$ does not have to be flat.

The angular momentum of $P$ corresponds to a non-zero PQ charge. The PQ charge is partially transferred into particle-antiparticle asymmetry of SM particles via the coupling of the axion with them and the SM interactions. The asymmetry is converted into baryon asymmetry via the electroweak sphaleron process. At the equilibrium, the baryon asymmetry $n_B$ normalized by the entropy density $s$ is given by
\begin{align}
\frac{n_B}{s} \equiv Y_B = c_B \left(\frac{T_{\rm EW}}{f_a}\right)^2 Y_{\rm PQ},
\end{align}
where $T_{\rm EW} \simeq 130$ GeV~\cite{DOnofrio:2014rug} is the temperature below which the electroweak sphaleron process becomes ineffective, $Y_{\rm PQ} = \dot{\theta} f_a^2/s$  is the PQ charge density normalized by the entropy density, and $c_B$ is a model-dependent constant given by~\cite{Co:2020xlh}%
\footnote{The signs of the first two terms are opposite to that in~\cite{Co:2020xlh}. This is because the sign of the axion-gauge boson coupling in~\cite{Co:2020xlh} is opposite to that used in~\cite{GrillidiCortona:2015jxo} and the literature on astrophobic axions, which stems from the sign convention of the Levi-Civita tensor.}
\begin{align}
    c_B \simeq & -\frac{21}{158}  + \frac{12}{79} c_W+ \sum_i\left( \frac{18}{79} c_{q_i} - \frac{21}{158}c_{\bar{u}_i} -  \frac{15}{158}c_{\bar{d}_i} + \frac{25}{237} c_{\ell_i}- \frac{11}{237}c_{\bar{e}_i} \right) \nonumber \\
    = &  \frac{-18+45(Q_{\bar{s}} + Q_{\bar{b}} )+63Q_{\bar{c}}  +22Q_{\bar{\tau}}}{1422} + \frac{12}{79}c_W.
\end{align}
In the second equality, we have imposed the astrophobic conditions.

The kinetic energy of the axion rotation is transferred into axion DM density, which is called the kinetic misalignment mechanism~\cite{Co:2019jts}. The number density of the axion $n_a$ normalized by the entropy density is   
\begin{align}
\frac{n_a}{s} \equiv Y_{\rm a} = c_{\rm DM} Y_{\rm PQ},
\end{align}
where $c_{\rm DM}$ is an $\mathcal{O}(1)$ constant. In the regime where axion DM is produced as a coherent oscillation of the axion field, which corresponds to $f_a \gtrsim 10^{10}$~GeV~\cite{Eroncel:2022vjg}, numerical and analytical computations show that $c_{\rm DM} \simeq 2$~\cite{Co:2019jts,Eroncel:2022vjg}. For lower decay constants, axion DM is produced via parametric resonance~\cite{Co:2021rhi,Eroncel:2022vjg}, and the precise value of $c_{\rm DM}$ is unknown. In this paper, we take $c_{\rm DM}$ as an unknown $\mathcal{O}(1)$ constant, anticipating that it will be determined by numerical computation in near future.

Requiring that the observed baryon asymmetry be explained by minimal axiogenesis and axion DM be not overproduced by kinetic misalignment, we obtain an upper bound on the decay constant, 
\begin{align}
f_a \leq 2.8 \times 10^6~{\rm GeV} \frac{c_B}{0.2} \frac{1}{c_{\rm DM}}.
\end{align}
The observed DM density is also explained when the inequality is saturated.

The upper bound is not compatible with the KSVZ and DFSZ axion models, which are subject to the  astrophysical lower bounds of $f_a> 10^{8-9}$ GeV. To overcome this difficulty, extra baryon or lepton number violations are introduced in~\cite{Co:2020jtv,Harigaya:2021txz,Chakraborty:2021fkp,Kawamura:2021xpu,Co:2021qgl,Barnes:2022ren}. Production of helical hypercharge gauge fields by the tachyonic instability induced by the axion velocity can produce baryon asymmetry without introducing extra interactions~\cite{Co:2022kul}, but fine-tuning of parameters is required.

The astrophobic axion model with $E/N=2$ may be consistent with the upper bound from minimal axiogenesis.
Unfortunately, the minimal model has $c_B \simeq 0.01$ and is not compatible with minimal axiogenesis. 
Non-minimal models may be compatible.
For example, for $(Q_{\bar{s}},Q_{\bar{b}},Q_{\bar{c}},Q_{\bar{\tau}},c_W)$= $(1,-2,-3,-2,0)$ and $(1,-1,-2,0,-2)$, $c_B\simeq 0.3$ and $0.4$, so $f_a \simeq 4$ and $ 6 \times 10^6 / c_{\rm DM}$ GeV is predicted, respectively. If $c_{\rm DM}\simeq 1$, the lower bound on $f_a$ from the cooling of neutron stars is satisfied for $z$ sufficiently close to $0.49$.

We note that the QCD axion mass predicted by minimal axiogenesis is around 1 eV, which is above the range of masses to which future IAXO helioscope~\cite{IAXO:2019mpb} will be sensitive. On the other hand, this is in a range in which optical haloscopes that search for absorption of DM have the best sensitivity to the axion-photon coupling~\cite{Baryakhtar:2018doz}, so we expect that experiments such as LAMPOST~\cite{Chiles:2021gxk} will probe this scenario. Moreover, in this range of axion masses, the axion-electron coupling required to explain stellar-cooling anomalies may also be within the reach of experiments aiming to detect DM via its absorption by molecules~\cite{Arvanitaki:2017nhi}. 

\section{Summary and discussion}
\label{sec:summary}
In this paper, we presented an astrophobic axion model where the couplings of the axion with nucleons, electrons, and muons are naturally suppressed. The axion decay constant $f_a$ may be as low as $10^7$ GeV. It is also possible to suppress the coupling with the photon so that the decay constant is even as small as $10^6$ GeV.

We studied the constraint from flavor-violating axion couplings. If the PQ charge of the strange quark is not the same as that of the down quark, the constraint from kaon decay requires special flavor structure, although not fine-tuned. If the PQ charges of the strange and down quarks are the same, generic flavor symmetry that explains the small down and strange Yukawa couplings significantly suppresses the axion-down-strange coupling and the constraint from kaon decay is avoided.

The astrophobic axion may have $f_a$ much below $10^{10-12}$ GeV, for which the axion abundance produced by the misalignment mechanism~\cite{Preskill:1982cy, Dine:1982ah,Abbott:1982af} or the decay of cosmic strings~\cite{Davis:1986xc} is much below the observed DM abundance. In addition to the kinetic misalignment mechanism discussed in Sec.~\ref{sec:axiogenesis}, decay of long-lived domain walls~\cite{Sikivie:1982qv,Chang:1998tb,Hiramatsu:2010yn,Hiramatsu:2012sc,Kawasaki:2014sqa,Ringwald:2015dsf,Harigaya:2018ooc} or parametric resonance~\cite{Co:2017mop,Harigaya:2019qnl,Co:2020dya,Nakayama:2021avl} can explain the observed DM abundance by axions.

The model can explain the hints for anomalous stellar cooling. The small axion-electron coupling required for the cooling can be obtained by a radiative correction from the coupling of the axion with other SM particles or by tree-level mixing of the electron or Higgs with PQ-charged heavy particles. For the former case, the decay constant needs to be below $\mathcal{O}(10^{6-7})$ GeV.  For the latter, the decay constant may be larger. Unless the (effective) PQ charge of the muon is the same as that of the electron, the axion generically has a flavor-violating coupling with the electron and muon that can be probed by $\mu\rightarrow e a$.

The baryon asymmetry of the universe may be explained by minimal axiogenesis. 
Unless some of the axion couplings are much larger than naive $1/f_a$-suppressed ones, the decay constant should be below $10^7$ GeV. This requires the suppression of the axion-photon coupling, which can be achieved for $E/N=2$.
The axion has a mass above eV and can be detected via absorption of axion DM. Since the axion-photon coupling is fixed up to the dependence on the axion mass, this scenario may serve as a benchmark for experiments such as LAMPOST that search for absorption of DM. Further requiring that the anomalous stellar cooling be explained, the axion should have a sizable coupling to the electron, which helps detection.

\section*{Acknowledgement}
The work of MB was partially supported by the National Science Centre, Poland, under research grant no.~2020/38/E/ST2/00243.

\bibliographystyle{JHEP}
\bibliography{refs}

\end{document}